\documentclass[aps,prd,twocolumn,groupedaddress,nofootinbib,showpacs]{revtex4}%
\usepackage{graphicx}
\usepackage{amsmath}
\usepackage{amssymb}
\usepackage{multirow}
\usepackage{bm}
\usepackage{color}
\usepackage[hypertex]{hyperref}

\begin{document}
\title{$J/\psi$ Pair Hadroproduction at Next-to-Leading Order in Nonrelativistic-QCD at CMS}

\author{Li-Ping Sun$^{a}$}
\email{sunliping@bucea.edu.cn}

\affiliation{ {\footnotesize (a)~School of Science, Beijing
University of Civil Engineering and Architecture, Beijing, China}}
%\date{\today}

\begin{abstract}

We perform a complete study on the $J/\psi$ pair hadroproduction at
next-to-leading order (NLO) in the nonrelativstic-QCD (NRQCD)
framework with the pair of $c\bar{c}$ either in ${}^{3}S_1^{[1]}$ or
${}^{1}S_0^{[8]}$ fock state. It is found that the ${}^{1}S_0^{[8]}$
channel contribution at NLO  is essential. Our results indicate that
for the CMS, the NRQCD predictions can not describe the experimental
data at all, and the total cross section predicted by NRQCD is
smaller than the experimental data by an order of magnitude. So new
mechanisms are needed to understand the CMS data for $J/\psi$ pair
production.

\end{abstract}
 \maketitle

{\it Introduction.}---Nonrelativistic QCD (NRQCD)\cite{nrqcd} is
widely used in the study of heavy quarkonium physics. In this
framework, a quarkonium production process can be factorized as the
multiplication of short-distance coefficients (SDCs) and
long-distance NRQCD matrix elements (LDMEs). The SDCs can be
calculated perturbatively and the LDMEs are strongly ordered by the
relative velocity $v$ between the quark and anti-quark inside of the
quarkonium. This factorization has been applied in single quarkonium
production and tested by various
experiments\cite{Inc3,Inc4,Inc5,Inc6}.

Besides the single quarkonium production, multi-quarkonuim
production provides complementary to understand the quarkonium
production mechanism. At the LHC, the LHCb Collaboration in 2011
measured the $J/\psi$ pair production for the first time at the
center-of-mass energy $\sqrt{s}=7~\mathrm{TeV}$ with an integrated
luminosity of $35.2~\mathrm{pb}^{-1}$\cite{LHCb}. In 2013, the CMS
Collaboration further released the data of $J/\psi$ pair
production\cite{CMS} with a much larger transverse moment range,
providing a good platform for testing the validity of NRQCD in
quarkonium pair production. Besides, the ATLAS Collaboration also
gives the measurement of the $J/\psi$ pair production\cite{ATLAS},
and a large transverse momentum cut is imposed on both the $J/\psi$.
%, meanwhile the center-of-mass energy $\sqrt{s}=8~\mathrm{TeV}$, so the ATLAS detection is a more suitable platform compared to the LHCb and CMS.

In Refs.\cite{LO1,LO2,LO3}, the leading order (LO) in $\alpha_s$
calculation of $J/\psi$ pair production in the color singlet model
(CSM) is performed. Relativistic correction to the $J/\psi$ pair
production is carried out in Ref.\cite{RC}, which makes significant
improvement for diluting the discrepancy between the LO results and
the experimental data. Furthermore, partial next-to-leading order
($\mathrm{NLO}^{\star}$) correction for $J/\psi$ pair production is
calculated by Lansberg and Shao \cite{NLOstar,DoubleJpsi}. They
argued that the $\mathrm{NLO}^{\star}$ yield can approach the full
NLO result at large $p_T$, which is the transverse momentum of one
of the two $J/\psi$'s, and thus the $\mathrm{NLO}^{\star}$ results
give a more precise theoretical prediction than the LO results in
this region. The full NLO predictions for color singlet(CS) channel
are obtained in our previous work \cite{nlo3s11}. Besides, the
complete LO predictions within NRQCD are obtained by Kniehl and
He\cite{LOcomplete}. All the above works are performed in the single
parton scattering (SPS) mechanism. Contribution of double parton
scattering (DPS) is assessed in
Refs.\cite{DoubleJpsi,DPS1,DPS2,DPS3}, which is expected to be
important. Besides, the color evaporation model is also used to
interpret the production of $J/\psi$ pair\cite{CE1,CE2}. As
predictions for DPS and color evaporation model are highly
model-dependent, it is needed to have an accurate calculation for
SPS contribution before one can extract the DPS contribution.

In order to further study the multi-quarkonium production, it is
necessary to evaluate the $J/\psi$ pair production to NLO for more
channels, which includes ${}^{1}S_0^{[8]}$, ${}^{3}S_1^{[8]}$ and
${}^{3}P_J^{[8]}$. Because ${}^{1}S_0^{[8]}$ is found to give the
most important contribution for single $J/\psi$ production
\cite{1201.2675,1403.3612}, in this letter we focus on the
${}^{1}S_0^{[8]}$ channel and evaluate each $J/\psi$ in
${}^{3}S_1^{[1]}$ and ${}^{1}S_0^{[8]}$ fock states to the NLO. The
calculations of ${}^{3}S_1^{[8]}$ and ${}^{3}P_J^{[8]}$ channels
will be studied in the future. Comparing to the LO result, NLO
result can not only decrease theoretic uncertainties, but also open
new kinematic enhanced topologies, which will dominate at large
$p_T$. More precisely, we will find that the differential cross
section $d\sigma/dp_T^2$ at large $p_T$ behaves as $p_T^{-8}$ at LO,
while it behaves as $p_T^{-6}$ at NLO due to double parton
fragmentation contributions \cite{DPF}.

{\it Formalism.}---In NRQCD factorization, the cross section of
$J/\psi$ pair production at the LHC can be expressed as \cite{nrqcd}
%%%%%%%%%%%%%%%%%%%%%%%%%%%%%%%%%%%%%%%%%%%%%%%
\begin{eqnarray}
d\sigma_{p+p \to J/\psi+J/\psi}=\sum_{i,j,n_1,n_2}{\int}dx_1dx_2{f_{i/p}(x_1)}{f_{j/p}(x_2)} \nonumber\\
\times~{d\hat{\sigma}^{n_1,n_2}_{i,j}}\langle\mathcal{O}_{n_1}\rangle^{J/\psi}
\langle\mathcal{O}_{n_2}\rangle^{J/\psi}. \label{eq:factorization}
\end{eqnarray}
%%%%%%%%%%%%%%%%%%%%%%%%%%%%%%%%%%%%%%%%%%%%%%%
where $f_{i/p}(x_{1,2})$ are the parton distribution functions
(PDFs), $x_1$ and $x_2$ represent the momentum fraction of initial
state partons from the protons,
$\langle\mathcal{O}_{n}\rangle^{J/\psi}$ are LDMEs of $J/\psi$ with
$n = {}^{2S+1}L_J^{[c]}$ in the standard spectroscopic notation for
the quantum numbers of the produced intermediate heavy quark pairs,
and $d\hat{\sigma}$ are partonic short-distance coefficients. In
this letter we set either $n_1=n_2={}^{3}S_1^{[1]}$ or
$n_1=n_2={}^{1}S_0^{[8]}$ in Eq.~\eqref{eq:factorization}.

\begin{figure}[!hbtp]
\centering
\includegraphics[scale=0.50]{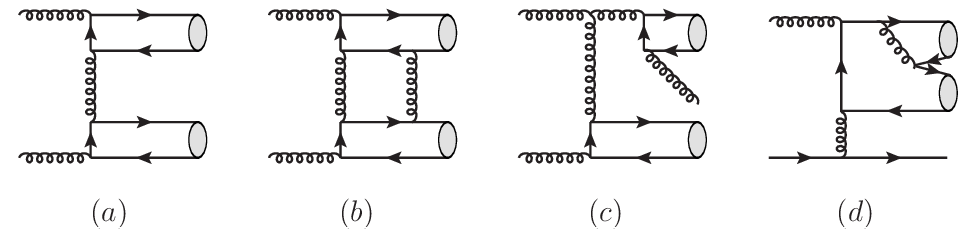}%
\caption{\small Typical Feynman diagrams for $J/\psi$ pair
production in ${}^{3}S_1^{[1]}$ and ${}^{1}S_0^{[8]}$ channels,
including LO and NLO.} \label{feynmandiag}
\end{figure}

In the LO calculation, there are two subprocesses:
$g+g{\rightarrow}J/\psi+J/\psi$ and
$q+\bar{q}{\rightarrow}J/\psi+J/\psi$, only the former of which is
taken into account since the contribution of the other process is
highly suppressed by the quark PDFs. While in the NLO case, besides
the gluon fusion process, the quark gluon process $q+g\rightarrow
2J/\psi+q$ should also be considered because they can give
non-negligible contribution. Typical Feynman diagrams at LO and NLO
are shown in Fig.\ref{feynmandiag}.

%%%%%%%%%%%%%%%%%%%%%%%%%%%%%%%%%%%%%%%%%%%%%%%%%%%%%%%%%%%%%%%%%%%
%%%%%%%%%%%%%%%%%%%%%%%%%%%%%%%%%%%%%%%%%%%%%%%%%%%%%%%%%%%%%%%%%%%

To tackle the infrared (IR) divergences in real corrections, the
two-cutoff phase space slicing method\cite{twocut} is employed.
After isolating the soft divergences and collinear divergences, the
cross sections for the $J/\psi$ pair production at the NLO can be
expressed as
\begin{eqnarray}
\sigma_{NLO}=\sigma_{Born}+\sigma_{Virtual}+\sigma_{Real}^{soft}+
\sigma_{Real}^{HC}+\sigma_{Real}^{\overline{HC}}, \label{eq:NLO}
\end{eqnarray}
where $HC$ and $\overline{HC}$ represent hard collinear and hard
non-collinear contributions, respectively. The soft divergences and
collinear divergences from real corrections will cancel divergences
from virtual corrections, and thus the final NLO contributions are
IR safe.

Because there are two $J/\psi$ states in the final state, the LO
contributions behave as $p_T^{-8}$ when $p_T$ is large. However, at
NLO level, there are contributions which give $p_T^{-6}$ behavior
\cite{DPF}[Fig.~\ref{feynmandiag} (c) and (d)]. We thus expect that
the NLO contribution will dominate at large $p_T$, especially for
the CMS and ATLAS data, where a relatively large lower $p_T$ cutoff
is taken\cite{CMS,ATLAS}. The expectation will be confirmed by our
numerical results shown below.

{\it Numerical~Inputs.}---Because of the complexity of the $J/\psi$
pair production, in our calculation, the package FEYNARTS
\cite{feynarts} is used to generate the Feynman diagrams and
amplitudes. The phase space integration is evaluated by employing
the package Vegas\cite{vegas}.

In numerical calculation, the CTEQ6L1 and CTEQ6M parton distribution
functions \cite{cteq1,cteq2} are used. The renormalization scale
$\mu_r$ and factorization scale $\mu_f$ are chosen as
$\mu_r=\mu_f=m_T$, with $m_T=\sqrt{p_T^2+16m_c^2}$ and charm quark
mass $m_c=M_{J/\psi}/2=1.55~\mathrm{GeV}$. In the two-cutoff method,
there are soft and collinear cutoffs, $\delta_s$ and $\delta_c$,
which we set to be $\delta_s=10^{-2}$ and $\delta_c=10^{-4}$.
Theoretical uncertainties are estimated by varying $\mu_r=\mu_f$
from $m_T/2$ to $2m_T$.

The CS LDME $\langle\mathcal{O}(^3\!S_1^{[1]})\rangle^{J/\psi}=1.16
\rm{GeV}^3$ is estimated by using the $\mathrm{B-T}$ potential
model\cite{BT}. While color octet(CO) LDME
$\langle\mathcal{O}(^1\!S_0^{[8]})\rangle^{J/\psi}=0.089
\rm{GeV}^3$, is taken from \cite{chao:2012}, which is determined by
fitting experimental data.

{\it Results.}---In the following, we give our results for the
$J/\psi$ pair production. In the CMS conditions~\cite{CMS}:
%%%%%%%%%%%%%%%%%%%%%%%%%%%%%%%%%%%%%%%%%%%%%%%%%%%%
\begin{eqnarray*}
&&|y(J/\psi)|<1.2~\mathrm{for}~p_T>6.5~\mathrm{GeV}, \mathrm{or}\nonumber\\
&&1.2<|y(J/\psi)|<1.43~\mathrm{for}~p_T>6.5\rightarrow
4.5~\mathrm{GeV},
\mathrm{or}\nonumber\\
&&1.43<|y(J/\psi)|<2.2~\mathrm{for}~p_T>4.5~\mathrm{GeV},
\label{eq:CMScut}
\end{eqnarray*}
%%%%%%%%%%%%%%%%%%%%%%%%%%%%%%%%%%%%%%%%%%%%%%%%%%%%
with $\sqrt{s}=7~\mathrm{TeV}$, the total cross section is measured
to be
%%%%%%%%%%%%%%%%%%%%%%%%%%%%%%%%%%%%%%%%%%%%%%%%%%%%
\begin{eqnarray}
\sigma_{Exp.}=1.49\pm0.07\pm0.14~\mathrm{nb}, \label{eq:CMSexp}
\end{eqnarray}
%%%%%%%%%%%%%%%%%%%%%%%%%%%%%%%%%%%%%%%%%%%%%%%%%%%%
while our LO and NLO calculations for the total cross section give
%%%%%%%%%%%%%%%%%%%%%%%%%%%%%%%%%%%%%%%%%%%%%%%%%%%%%%%%%%%%%%%%%%%
\begin{eqnarray}
\sigma_{\mathrm{LO}}=(0.048+0.014)~\pm0.02~\mathrm{nb},\nonumber\\
\sigma_{\mathrm{NLO}}=(0.18+0.03)\pm0.10~\mathrm{nb}.
\label{eq:CMSresults}
\end{eqnarray}
%%%%%%%%%%%%%%%%%%%%%%%%%%%%%%%%%%%%%%%%%%%%%%%%%%%%%%%%%%%%%%%%%%%
the first value in the bracket represents the CS contribution, while
the second one represents the CO contribution, and the uncertainties
come from the $\mu_r=\mu_f$ varying from $m_T/2$ to $2m_T$. As
expected, we find the NLO calculation gives the dominant
contribution. In (\ref{eq:CMSresults}) the contribution of feeddown
process $p+p\rightarrow J/\psi+\psi(2S)+X\rightarrow 2J/\psi+X$ and
$p+p\rightarrow J/\psi+\chi_{cJ}+X\rightarrow 2J/\psi+X$ are also
included, which are estimated to be $30\%$ of the direct
production\cite{LO3}. Comparing (\ref{eq:CMSexp}) with
(\ref{eq:CMSresults}), we can see the cross section measured by CMS
can not be described by the NRQCD calculation at NLO.

\begin{figure}[!hbtp]
\centering
\includegraphics[scale=0.65]{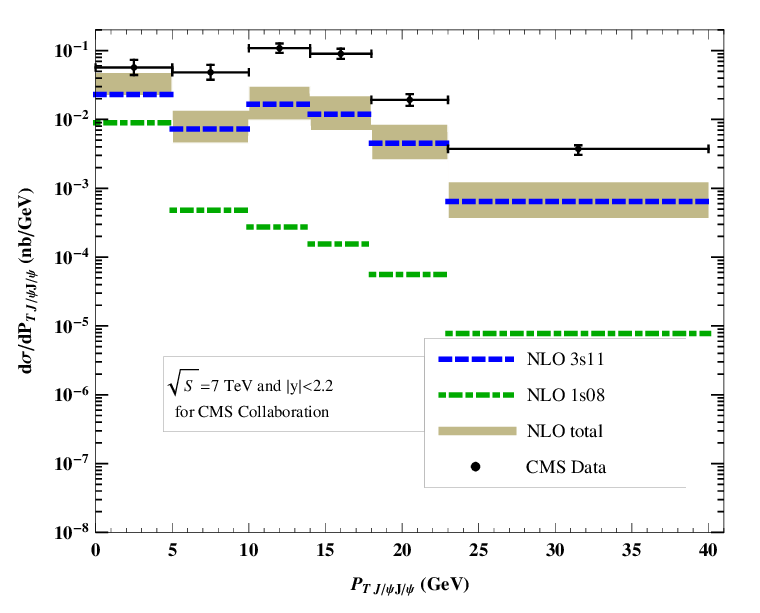}%
\caption{\small (color online). Differential cross sections in bins
of the transverse momentum of $J/\psi$ pair at CMS. The data are
taken from Ref.~\cite{CMS}, and the $\mathrm{NLO^{\star}}$ results
are taken from Ref.~\cite{NLOstar}. The dashed and dot dashed lines
denote the NLO ${}^{3}S_1^{[1]}$ and ${}^{1}S_0^{[8]}$ results
respectively, and the band denotes the NLO total result, where the
uncertainties are due to scale choices as mentioned in the text.}
\label{pjjCMS}
\end{figure}

We then compare our prediction for the transverse momentum $p_{T
J/\psi J/\psi}$ distribution of $J/\psi$ pair with the CMS data and
the $\mathrm{NLO^{\star}}$\cite{NLOstar} yields. The result is shown
in Fig.~\ref{pjjCMS}. At LO, $p_{T J/\psi J/\psi}$ is always zero,
because it is a two-body final state process.  At NLO, we can first
find that the contribution of the ${}^{1}S_0^{[8]}$ channel is small
even if at the large $p_{T J/\psi J/\psi}$, this is normal, because
we believe the dominant contribution at large $p_{T J/\psi J/\psi}$
may come from the ${}^{3}S_1^{[8]}$ channel, which is our next work.
we also find that the behavior of the NRQCD result is similar to the
experimental data, but smaller than the data by an order of
magnitude. For the $\mathrm{NLO^{\star}}$, it is consistent with our
NLO prediction at large $p_{T J/\psi J/\psi}$. The data obviously
overshoots our NLO prediction at the whole $p_{T J/\psi J/\psi}$
region. Because both CS contribution and dominant CO contribution
have been considered, we concluded that, the NRQCD factorization can
not describe the CMS data even after the NLO correction.  Therefore,
other mechanism must be included, besides the SPS contribution in
the NRQCD framework, to explain experimental data.

%%%%%%%%%%%%%%%%%%%%%%%%%%%%%%%%%%%%%%%%%%%%%%%%%%%%%%%%%%%%%%%%%%%
%%%%%%%%%%%%%%%%%%%%%%%%%%%%%%%%%%%%%%%%%%%%%%%%%%%%%%%%%%%%%%%%%%%

\begin{figure}[!hbtp]
\centering
\includegraphics[scale=0.65]{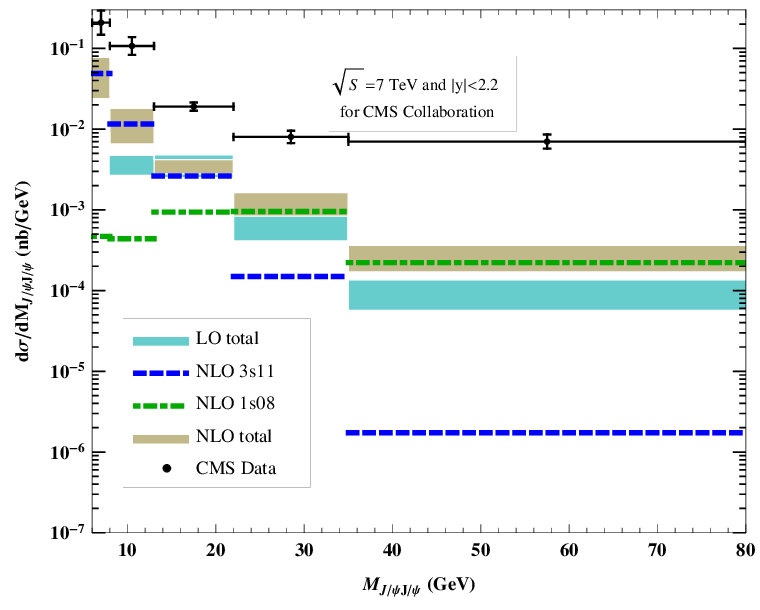}%
\caption{\small (color online). Differential cross sections in bins
of the $J/\psi$ pair invariant mass at CMS. The data are taken from
Ref.~\cite{CMS}, and the $\mathrm{NLO^{\star}}$ results are taken
from Ref.~\cite{NLOstar}. The dotted, dashed and dot dashed lines
denote the NLO ${}^{3}S_1^{[1]}$, LO ${}^{1}S_0^{[8]}$ and NLO
${}^{1}S_0^{[8]}$ results respectively, and the two bands denote the
LO and NLO total results, where the uncertainties are due to scale
choices as mentioned in the text.} \label{MassCMS}
\end{figure}

The invariant mass distribution (denoted as $M_{J/\psi J/\psi}$) for
CMS is shown in Fig.~\ref{MassCMS}. We can see that the
${}^{1}S_0^{[8]}$ channel has big contribution in medium and large
$M_{J/\psi J/\psi}$ region, which is more important comparing with
the ${}^{3}S_1^{[1]}$ channel. The sum of ${}^{3}S_1^{[1]}$ and
${}^{1}S_0^{[8]}$ channel again indicates that the NLO result can
not describe the CMS data. Like the $p_{T J/\psi J/\psi}$
distribution,  the NLO prediction for  the $M_{J/\psi J/\psi}$
distribution is smaller than the experimental data by at least one
order of magnitude for each bin, which also reflects the fact that,
in the $J/\psi$ pair production, the NRQCD prediction contributes
little.

%%%%%%%%%%%%%%%%%%%%%%%%%%%%%%%%%%%%%%%%%%%%%%%%%%%%%%%%%%%%%%%%%%%
%%%%%%%%%%%%%%%%%%%%%%%%%%%%%%%%%%%%%%%%%%%%%%%%%%%%%%%%%%%%%%%%%%%

\begin{figure}[!hbtp]
\centering
\includegraphics[scale=0.65]{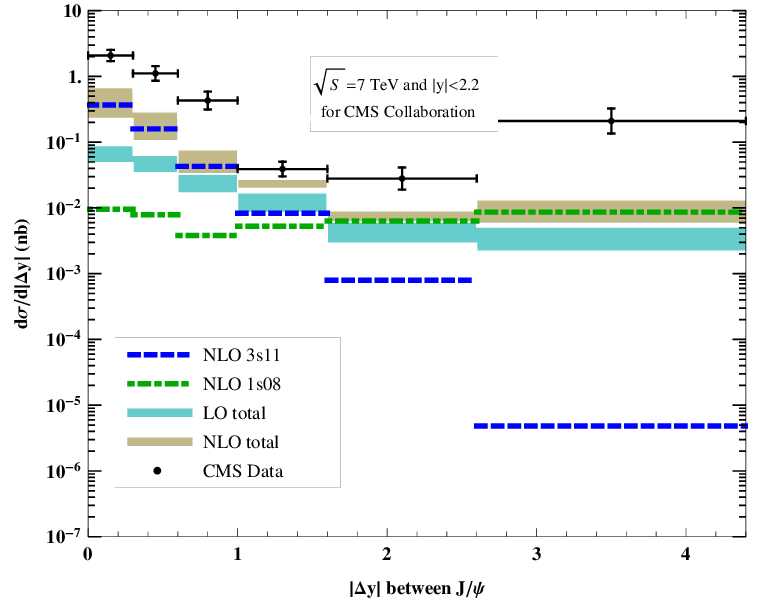}%
\caption{\small (color online). Differential cross sections in bins
of the $J/\psi$ pair $|\Delta y|$ at CMS. The data are taken from
Ref.~\cite{CMS}, and the $\mathrm{NLO^{\star}}$ results are taken
from Ref.~\cite{NLOstar}.. The dotted, dashed and dot dashed lines
denote the NLO ${}^{3}S_1^{[1]}$, LO ${}^{1}S_0^{[8]}$ and NLO
${}^{1}S_0^{[8]}$ results respectively, and the two bands denote the
LO and NLO total results, where the uncertainties are due to scale
choices as mentioned in the text.} \label{DyCMS}
\end{figure}

The $J/\psi$ pair rapidity difference $|\Delta y|$ distribution for
CMS is shown in Fig.~\ref{DyCMS}. We see that the ${}^{1}S_0^{[8]}$
channel also has big contribution in the medium and large $|\Delta
y|$ region, and at large $|\Delta y|$, the ${}^{1}S_0^{[8]}$ channel
is dominant. Even though, the sum of ${}^{3}S_1^{[1]}$ and
${}^{1}S_0^{[8]}$ channels can not describe the CMS data, similar to
the above two distributions.
%, in the
%$|\Delta y|$, the NLO prediction is also smaller than the
%experimental data by at least one order of magnitude for each bin.
%For the $\mathrm{NLO^{\star}}$ result, at small $|\Delta y|$ region,
%it is even closer to the CMS data than the NLO prediction, but at
%large $|\Delta y|$ region, the $\mathrm{NLO^{\star}}$ result
%decreases rapidly than the NLO behavior.

%%%%%%%%%%%%%%%%%%%%%%%%%%%%%%%%%%%%%%%%%%%%%%%%%%%%%%%%%%%%%%%%%%%
%%%%%%%%%%%%%%%%%%%%%%%%%%%%%%%%%%%%%%%%%%%%%%%%%%%%%%%%%%%%%%%%%%%

{\it Summary.}---In the framework of NRQCD factorization, we
evaluate the full NLO $J/\psi$ pair production via the
${}^{3}S_1^{[1]}$ and ${}^{1}S_0^{[8]}$ channels. We find that NLO
corrections are essential for $J/\psi$ pair production, compared to
the LO results. For the CMS, the NLO predictions of total cross
section, $p_{T J/\psi J/\psi}$ distribution, invariant mass
distribution of the $J/\psi$ pair, and rapidity difference
distribution of the $J/\psi$ pair are much smaller than CMS data by
about an order of magnitude.
%, the data obviously overshoot our NLO prediction at the whole $p_{T J/\psi J/\psi}$, $M_{J/\psi J/\psi}$ and $|\Delta y|$ region, so are the $\mathrm{NLO^{\star}}$ results.
This reveals the signal that, in the $J/\psi$ pair production
process, the NRQCD NLO result is not the dominant contribution,
there must be some new schemes dominating the process, if the CMS
data are confirmed.

%%%%%%%%%%%%%%%%%%%%%%%%%%%%%%%%%%%%%%%%%%%%%%%%%%%%%%%%%%%%%%%%%%%%%
We thank Y. Q. Ma and C. Meng for valuable discussions and
suggestions. This work was supported by the National Natural Science
Foundation of China(NSFC) under grants 11905006.

%%%%%%%%%%%%%%%%%%%%%%%%%%%%%%%%%%%%%%%%%%%%%%%%%%%%%%%%%%%%%%%%%%%%%%
%\vspace{1.5cm}

\end{document}